# A system outlook on the vision problem associated with observation of light flickering at the micro-saccades' frequency

*Emanuel Gluskin[1,2,3,4] ✻, Yehuda Ben-Shimol[5], Frangiskos V. Topalis[6], Nikolas Bisketzis[6].*

[1]The Galilean Sea Academic College, [2]Ort Braude Academic College, [3]HIT Holon,

[4]EE BGU, Beer-Sheva 84105. Israel, (http://www.ee.bgu.ac.il/~gluskin/)

[5]Department of Communication Systems Engineering, Ben-Gurion University, Beer Sheva, 84015,
benshimo@bgumail.bgu.ac.il
[6]National Technical University of Athens, School of Electrical and Computer Engineering, 9 Iroon Politechniou St., 15780 Athens, Greece.  topalis@softlab.ntua.gr, bisketzi@mail.ntua.gr

✻**Corresponding author**:  email: **gluskin@ee.bgu.ac.il**

**Abstract:** The flickering of the light of fluorescent lamps (FL), whose basic frequency, 100 Hz, is close to that of the micro-saccadic (the eye-muscles' tremor component) eye movement, is a severe problem for autists (autistic humans), a problem for newborn babies, for people after some traumatic accidents, and for some 10% of otherwise absolutely normal humans. Taking the line of a "system-terms" discussion of the vision-disturbance problem, the present work provides a constructive framework for investigating the problem.  Using the results of light intensity measurements and some simple analytical models for the instantaneous light-intensity function $\psi(t)$, and analyzing the role of the coincidence (closeness) of the frequencies of the ripple of $\psi(t)$ and the micro saccades, we suggest a block-diagram for the biological vision control system.  We also show that a singularity of the waveform of $\psi(t)$, which is typical for most FL, may be important for brain control of the eye tremor.  The latter conclusion is supported by a simulation of the image intensity on retina.

# 1. Introduction

## 1.1. The general problem

Most artificial (lamp) lighting is fluorescent. The problem in focus, associated with this actual situation, is that about 50% of autists (autistic humans) suffer [1] from fluorescent light whose flickering is not (at least not consciously) seen by most people. The vision problem is inherently relevant, to a degree, to at least some 10% of normal population [1,2], and also to some people after traumatic incidents. Newborn babies are also sensitive to fluorescent light, which attracts their sight.

The problem is not with the light spectrum, which is quite good in any kind of modern lighting, but with *the flickering of the light at twice the frequency of the line voltage, which is rather close to the frequency of eye-muscles tremor (the micro saccades).*

This frequency correlation leads to some objective problem with the eye-control, relevant to all humans. The possible *difficulty in performing* the required eye-control may be, however, individual for each of the above-mentioned specific groups, in particular, for autists, for whom failing in performing the control leads to a behavioral problem. The anxiety reaction of the autists, which motivated this research, may be compared with the reaction of mentally sick people, in the known psychoanalytic investigations by Sigmund Freud, whose conclusions relate also to normal humans.

According to the classical works [22,23], the degree of the unconscious stressing of the eye muscles (associated with a conditioned reflex) completes the conscious visual perception of the object in the sense of the estimation, by the brain, of the size of (or, the distance to) the object. Basic information about eye and vision in general, may be found in [3-17].

Analyzing the dynamic eye-control problem associated with the flickering, we suggest some system modeling, and present some numerical simulations supporting the formulae analysis. The work thus tries to introduce of some new tools for the analysis of the vision problem, which may attract the attention of professional system specialists, whose understanding of what should be a system definition for an analytical study and of how to systematically develop the system analysis could contribute to biological and physiological studies. Our conclusions regarding the control of the eye muscles' tremor may also be relevant to studies of basic conditioned reflexes, but in a dysfunctional context. In particular, our analysis suggests that for the light flickering at the problematic frequency, the phase of the tremor should be controlled by the brain.

### *1.2. The specific autistic problem*

Though we start the analysis in Section 3 from a general vision problem, the humanitarian aspect of the autism-study [1] requires some additional information to be kept in one's mind. In general, when speaking about autistic vision problem, one is occupied both with the physical and behavior problems.

Just to give some brief background of the behavioral problem, let us note that the methods of reducing the flicker in the total illumination, given by several differently fed lamps, sufficient for one who works with machines having open rotating parts (that may be seen in the stroboscopic light as stationary), are partial solutions for autists for whom *the very source* of *the light* may be the object for observation and worry. The sight of an autist may become "locked" on to the flickering lamp [1,18-21], and each of the "mutually compensating" lamps may still be a problem. Even if for a lamp the ripple of the light is not strong, it can be detected by autists by focusing their eyes on the lamp. Some architectural solutions may be needed for 'hiding' the lamps from direct sight, but usually these are not provided.

Among fluorescent lamps (below 'FL'), only those that are fed at high (20 kHz – 50 kHz) frequencies by means of electronic ballasts ("gears") directly give non-flickering light. Such arrangements and fixtures should be used in autists' schools and houses ([19,2] for a survey of lighting methods), but will not be met at every place where autists can find themselves. Since autists often are guided, it would be good if a special personal "vision chip" predicting the response of a certain autist to actual lighting situation could be developed, helping the parents, teachers, and tutors.

The physical and biological structure of the eye of an autist is normal, and the problem is only in the eye control and the associated information processing by the visual cortex that is presumably influenced by some excessive brain noise. However, the physical normalness of the eye does not mean, for instance, that the typical angle-limits of the micro saccades cannot be increased when an autist has the vision problem. Problematicity in control of the eye may make one anxious and influence the eye movement in different ways especially because the usual slow saccades and the micro saccades are done by the same eye muscles.

In any case, our focus since now is not on any humanitarian "ergonomic" aspects, but possible contribution to the study of the brain activity as far as it can be associated with the specific vision problem.

### 1.3.  The dynamic eye-control problem

It is well known that an absolutely static image on the retina quickly ceases to be registered, and saccadic eye movement is necessary for observing stationary objects, here the lamp.  The saccadic movement has a slow 3–5 Hz component, which is (unconsciously for us) controlled by the brain, and some 100 Hz micro-saccadic tremor caused by the eye muscles.  After explaining how the coincidence of the basic frequency of the ripple of the light with that of the tremor influences the "input perception" of the vision process, we formulate a reason for the tremor to be controlled by the visual cortex.  This control should not be a difficult problem in normal humans, but may be very problematic in autists.

   The information needed regarding fluorescent lighting per se is taken from [18-21].  Besides the immediately important fact that the basic frequency of the light intensity function (or its ripple) of many FL is that of (or very close to) the micro saccadic tremor of the eye muscles, the additional interesting point is that the light-intensity function is usually singular, having a discontinuous derivative.  This should be relevant for correlation effect obtained at any frequency.

## 2.  The light-intensity function $\psi(t)$

The *instantaneous light intensity function* $\psi(t)$ had been measured (in the Photometry Laboratory of National Technical University of Athens) electronically at a small distance from the lamp.  It was found most suitable to measure $\psi(t)$, using a simple photodiode; the quick processes in the photodiode make the measurements to be reliable regarding obtaining the correct waveform of $\psi(t)$.

   Using the theory of [21], the following formula is obtained in [20]:

$$\psi(t) = \psi_{\min} + D|i(t)|, \tag{1}$$

where $i(t)$ is the lamp's current function, and $D$ is a constant.  The relative depth of the "modulation" of $\psi(t)$, i.e. $Di_{max}/\psi_{\min}$, may be up to 20% or even 30%.

   The experimental $\psi(t)$ of a very common-type fluorescent lamp is shown in Fig. 1 to be in close agreement with (1).

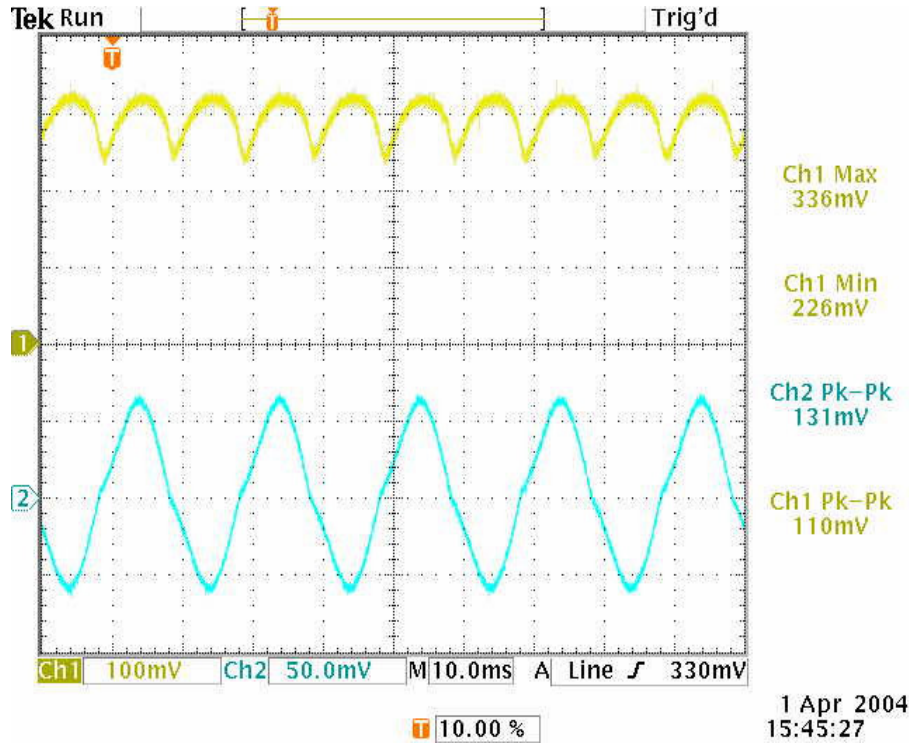

<u>Figure 1</u> - The $\psi(t)$ (upper wave) and $i(t)$ for a T-8 (28 mm diameter), 36 W FL, 50 Hz. Compare the basic frequencies of the two waves.

It needs to be stressed that the basic harmonic of $|i(t)|$ is *double* of that of $i(t)$, i.e. 100 Hz in Europe. This also follows from the fact that the basic frequency of the light ripple must equal that of the instantaneous power function $p(t) = v(t)i(t)$ of the lamp, while the latter is 100 Hz for any model of the lamp's voltage-current characteristic. (Also for incandescent lamp, the basic frequency of $p(t)$ is 100 Hz, but for incandescent lamp the very significant thermal inertia of the filament does not allow the ripple of $p(t)$ to cause any noticeable ripple of $\psi(t)$; the physics of the fluorescent lamp [18,21] is associated with much quicker processes.)

Since the current wave does not strongly differ from a sinusoid, replacement of (1) by

$$\psi(t) = \psi_{\min} + Di_{\max} |\sin \omega t| \ , \qquad \omega = 2\pi 50 \text{ Hz} \ ,$$

is not associated with any rough mistake; while, as the point, the singularity of the true $\psi(t)$ is well expressed.

Below we consider both the singular and the purely sinusoidal (of the frequency 100 Hz) ripple that may be obtained by taking only the basic harmonic of the any ripple, which gives

$$\psi(t) = \Psi + B \sin 2\omega t \qquad (2)$$

with constants $\Psi > \psi_{min}$ ($\Psi$ is the average of $\psi(t)$) and $B$. For some fluorescent lamps of old technology [18], model (2) may be satisfactorily, especially for a lamp with relatively large mass of the gas, which depends on the gas pressure and the volume of the tube.

We shall show that despite the same basic period, the cases of (1) and (2) are distinct for the eye perception, i.e. the singularity of $\psi(t)$ is important.

The consideration, near the source, of $\psi$ as a function of $t$ only, is sufficient for the study of light flickering per se. When analyzing the visual response, we have to also use the multi-argument function $\psi(\mathbf{r}, t)$ where $\mathbf{r}$ is a radius-vector, related to a spatial point. Finally, we are focused on variations of the light intensity at certain, arbitrary chosen, point on the retina, and $\mathbf{r}$ here is a point in a plane.

We come to the main point.

## 3. The role of the coincidence of the frequency of the ripple of $\psi(t)$ and the frequency of the micro saccades

The saccadic eye movement ensures that *the optical image* of an object stationary in space (as the FL is) *moves on the retina*, and *thus* an area of the retina's elements is (re)activated by the photo-excited chemical processes. On the background of the physiological necessity of the movement of the image on retina, the fact of interfering influence of the ripple of $\psi(t)$ on vision suggest introducing a time derivative of the signal to the analysis as follows. *We introduce time-derivative of the* (*moving*, *because the retinal is moving*) *information field on the retina, taken along a path of some scanning of this field by edges of the optic nerve that connects the retina (eye) to the visual cortex*. The associated use of both $\psi(t)$ and $d\psi(t)/dt$ (or, now, $\partial\psi(\mathbf{r}, t)/\partial t$) also agrees with the natural mathematical requirement that when the eye movement is considered, vision becomes a dynamic process whose state-space should possess a proper dimension.

The retina's surface is denoted as $(x, y)$. The optical (purely geometrical, at this stage) image of FL (a spatial function that is denoted below as $\zeta(\cdot)$) moves on $(x, y)$ because of the tremor. Since only *mutual movement* of the retina and the image is important, it is also possible to define $(x, y)$ not as the retina's but as the "image's surface", that is, the surface in which the image is unmoving. Then the edges of the optic nerve connected to the retina are seen as "slipping" on this unmoved surface, performing the information "scanning".

The general mapping action of the crystalline eye lens is $\psi(\mathbf{r}, t) \rightarrow \phi(x, y, t)$, or $\psi(\mathbf{r}, t) \rightarrow \mathbf{A}(x, y, t)$, with the respectively obtained *scalar* ($\phi$) or *vector* ($\mathbf{A}$) informative field (the "image") on retina. We mostly use $\phi(x, y, t)$, but, in principle, both scalar and vector cases are treated similarly. For $\phi(x, y, t)$, the full time derivative is

$$\frac{d\varphi}{dt} = (grad\,\varphi)\,\boldsymbol{v} + \frac{\partial\varphi}{\partial t}, \tag{3}$$

and for $\mathbf{A}(x, y, t)$ it is

$$\frac{d\mathbf{A}}{dt} = (\mathbf{v} \cdot \nabla)\mathbf{A} + \frac{\partial\mathbf{A}}{\partial t}, \tag{4}$$

where vector $\mathbf{v}$ is the velocity of reading the information along the 'trajectory' of the scanning, i.e. the velocity of the mutual movement of the retina and the image.

It is sufficient to consider only one trajectory of the scanning associated with one edge of the vision nerve. Each edge (receiving information [9-17] from more than 150 among the $10^6$ receptors of the retina) gives such a trajectory, because of the tremor.

The difficulty in distinguishing, while reading the information, between the two different terms in the same expression for $d\phi(t)/dt$ (or $d\mathbf{A}/dt$), appearing in the right-hand side of (3) or (4), may result in difficulty in distinguishing between the light intensity variations, represented by $\partial\psi/\partial t$ and the movement of the associated object (which may be non-flickering) in the space, in processing the information.

Remark:   Observe that the very saccadic movement aimed to reactivate the chemical processes, uses the equivalence of the two terms of the full time derivative. The chemical processes at a point on retina are local, and for the individual receptors the movement of the light spot can be equivalent to flickering.

In the dependence of **v** on time, the frequency of the eye muscles' tremor is influential, while it is important for the use of (3,4) that the coincidence of the frequencies of **v** and $\phi$ (or $A$) means (is caused by) the coincidence of the tremor's frequency with the frequency of $\psi(t)$ .

### 3.1. The "resonant pumping" of the visual information and some time averaging of the signal at the inputs of the fibers of the optic nerve

Since the excitation of the receptors implies some energy consumption, function $\phi$ is essentially positive, and the coincidence of the frequencies of **v** and $\phi$ displays some analogy with familiar resonant processes where an increase of the amplitude, i.e. an *energy 'pumping',* is typical. In the present section, we treat the meaning of this analogy in system terms; more physical aspects are considered in Sections 3.2 and 9.

Let us note that any resonant output essentially depends on the phase shift between the input and the response functions. In simple resonant systems, the input directly causes the output, and there thus is an automatic adjustment of the phase shift, so that the "scalar product" of the input and the response is large, i.e. the energy pumping is effective. (Think, e.g., that you are pushing a swing carrying a child.) In our case, however, the tremor exists anyway, and, initially, its phase is not adjusted to that of $\psi(t)$ ; it should be adjusted by the eye-control system.

Generally, in resonant processes, the final (established) value of the amplitude is defined not only by the intensity of the energy supply but by equilibrium between the energy supply and the dissipation. For the eye, the decaying processes in the retina can play the role of the dissipation. In the present work, we cannot analyze these decay processes. Our point is to consider how the brain can prevent the pumping of the visual information by means of phase control, i.e. by weakening the very resonant excitation, or, in physiological terms, the effect of the prolonged fixation associated with constant illumination at some points.

While speaking below about the average value of the signal at the input of a fiber of the optic nerve, we consider, using (3) or (4), the intensities of the *time averages of the "scanning response"*, $d\phi / dt$ (or $d\mathbf{A} / dt$ ). This average is interpreted here as the intensity of the resonant "pumping". As any averaging, this averaging must be associated with some integral-type effect, in this case a physiological one.

Since the time-average of the *partial* derivative by time of a periodic by '*t*' function is zero, the average of the left-hand side of (3) or (4) is given by the value (everywhere below '$\langle \ \rangle$' means '$\langle \ \rangle_t$', i.e. averaging *by time*):

$$\langle \mathbf{v} \cdot \mathrm{grad}\, \phi \rangle \quad (\text{or } \langle (\mathbf{v} \cdot \nabla)\mathbf{A} \rangle) , \tag{5}$$

which is, respectively, a scalar or a plane-vector. If the time-shift between $\mathbf{v}$ and $\phi$ (or $\mathbf{A}$) does not provide a small value of (5), then the average $\langle d\phi / dt \rangle$ (or $\langle d\mathbf{A} / dt \rangle$) is not small too, and the scanning results, at least in some points, in some integral "pumping" of an informational measure of the signal into the cortex. This should be considered as causing difficulty in processing the signal/image. (See also Section 9).

The average (5) may be small if the phase of the tremor is quickly (or randomly) changed, or if the phase is changed slowly, *but properly*. The first case is rejected by the certainty (not ideal, of course, but the tremor is not a random process) of the frequency of the tremor. In the second case we speak about *minimization* of (5) by the vision cortex , assuming that the cortex *must control* the 100 Hz micro saccades.

*We interpret the vision problem, most sharply presented in autusts, as the difficulty of the cortex, -- that is also influenced by the abnormal autistic brain activity, – in controlling the phase of the tremor minimizing the average* (5), *i.e. the information pumping to the cortex, caused by the light flickering at 100 Hz, i.e. at the frequency of the micro saccades.*

It needs to be stressed that for any certain phase shift between the tremor and the ripple, the average that we consider depends on the "trajectory" of the certain point on retina, and is different for different points of the retina. That is, in the synchronous-movement case, the intensity of the light at some groups of the points of the retina will (all the time) be stronger than at some other groups. Since the *overall* (both temporal and spatial, over the whole illuminated area) average of the intensity is given by the total input light input (i.e. is fixed), and the total area, the time-average *of the signal* (i.e. of $\phi$ , or $\mathbf{A}$, by itself, not of the derivative) at any certain group of the retina sensors (or over certain fiber of the optical nerve) cannot be made arbitrary small. In fact, we have to focus at the unevenness of the distribution of the time-averaged intensity over different fibers of the optic nerve. We assume, in particular (see also Section 6) that at the places where the averaged intensity is maximal, the vision has a special problem.

### *3.2. The physical side*

Our almost purely geometrical approach to the complicated physiological structure requires some physical reservations, one of which is given now, and some others in Section 9.

For the retinal perception, i.e. for the input signals of the fibers of the optic nerve, it is not the light intensity (i.e. the incident flow of the photons) which is directly relevant, but the products/state of the resulting photo-excited chemical process. (See also Section 9 for this point.) Thus, in addition to the concept of "information field", the concept of "vision-sensibility field" should be used. The considered time-averaging is, in more precise terms, *a low-frequency filtration*. Taking the time-average of a periodic process means an *ideal* low-frequency filtration at a point of the retina, where only the dc component is allowed to pass via an ideal filter. Certainly, we do not have any such ideal filter in the biological system, and the averaging may be performed only partly by means of some integration-type process. Thus some partial transfer of the 100Hz to the brain is inevitable.

Assuming that the chemical response may be seen as a linear time-variable system, we can write the physiological input signal at the inputs of the optical nerve fibers as

$$\text{(The physiological input of the optical nerve)}(t, \boldsymbol{r}) \; = \; \int_0^\infty h(t, \lambda) \rho(\lambda, r) d\lambda$$

where $h$ is the intensity of the chemical response, and $\rho(t, \boldsymbol{r})$ is, e.g. of the type (7). If we expect $\rho(t, \boldsymbol{r})$ to be periodic by $t$, the above function should also be periodic by $t$, at every point (this is physically provided, in particular, by the usual saccadic eye movement); one may then try to use a linear time-invariant model for the chemical response, with $h(t-\lambda)$ in the integrand.

However, even if being non-ideal, the averaging undoubtedly makes the phase shift between the tremor and the ripple an important parameter, and the concrete results of Section 6 will confirm that the latter is a very useful point of view.

## 4. A proposition for analysis of the biological vision control in system terms

The present state of the knowledge regarding human physiological vision is very poor, and there is no direct possibility to check by a reliable measurement what image on retina the brain prefers. Certainly, it is still impossible to speak about the human physiological vision system as about some *definable* system, in the sense accepted in the theory of electrical systems. It is important thus to suggest some simple models for the analysis of the problem, helping to introduce some description terminology, outlining points for constructive analysis, and hopefully attracting system-specialists to this important research field.

Figure 2 suggests a control loop for modeling the visual cortex operation. This loop involves the special block "Processing – 2" which interacts with a more general-propose block "Processing - 1" that receives information of any kind from the eye, e.g., about other objects in the scope of vision, and usually receives the 100 Hz signal of the tremor by the reasons discussed in Section 8.1. All of the channels coming to "Processing – 1" and "Processing – 2" represent the action of the optic nerve; $F(\cdot)$ is here the nerve-pulse form of the information.

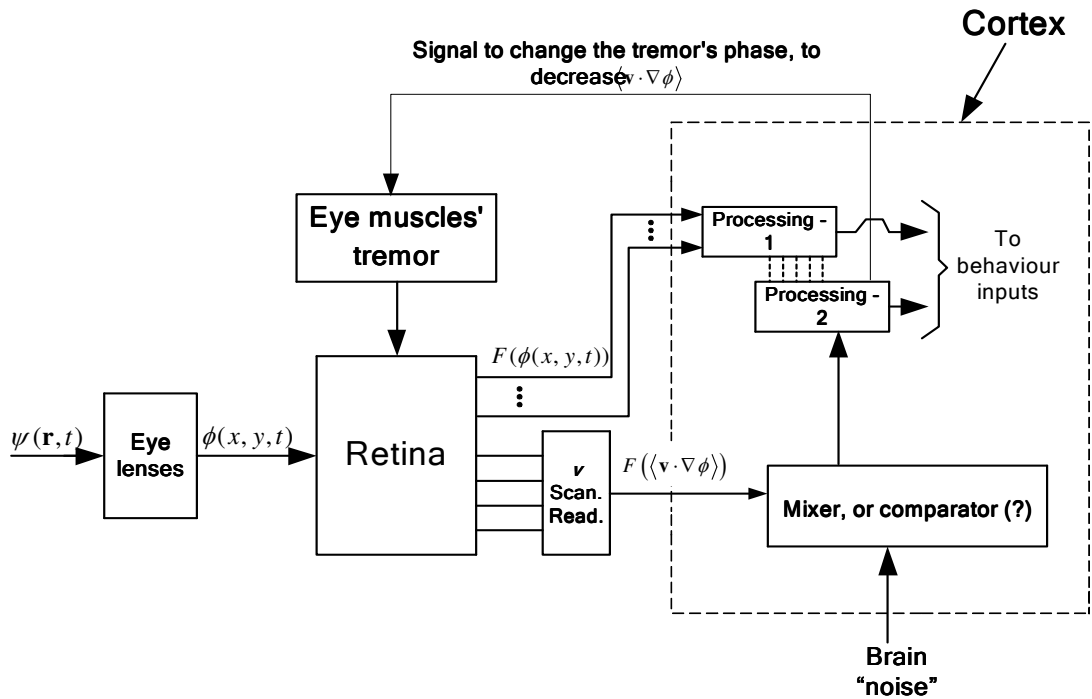

Figure 2 - The schematic eye-cortex feedback connection minimizing the time-average (5) at a point where it is maximal. Here $\phi(\cdot,\cdot,\cdot)$ is the information distributed on the retina, and $F(\cdot)$ is the nerve-impulse form of the scanned signal which is being sent to the visual cortex. The averaging of the scanned signal in time, associated with the coincidence of the frequencies of $\psi(t)$ and the muscles' tremor, which is seen as important, motivated this scheme. The system tries to make $\langle \mathbf{v} \cdot \nabla \varphi \rangle$ small, but can do this only when $F\left(\langle \mathbf{v} \cdot \nabla \varphi \rangle\right)$ is larger than the autistic "brain noise".

In terms of this control scheme, the essence of the autistic vision problem is that when (5) is provided by the control to be small, it, *as the input of the loop*, "disappears" in the brain's "noise" that influences that vision cortex; the control is *lost*, and (5) becomes increasing at some ports.  This requires control to be established again, which takes, however, some time, which might start to cause the autist's anxiety.

 The threshold for the control to exist, is defined by the level of the brain's "noise", while the intensity of the flickering which starts to cause the vision disturbance and worry in an autist is a quantitative measure of the "noise", -- a good point for a biological experiment.

 This scheme for the treatment of the vision problem can be constructive for the development of "autistic" *vision chips* [24-33] in which the already developed methods for modeling retinal and vision processing would act via elements of such a control loop.  Our opinion is that the artificial vision science may thus contribute to the study of the vision disturbance problem.

 A suggested experiment for checking the scheme of Fig. 2 might be to try to obtain a quiet state in an autist, in which the brain-noise would, presumably, be reduced.  Our model suggests that in the quiet state the irritation caused in the autist by a fluorescent lamp would be reduced.

### 5.   A simplification of the analysis: focusing on the optical image

Not contradicting anything in the above general outlook, let us now be focused just on the patterns of the image created on retina.

 The situation re the image is very close to that of obtaining a stroboscopic picture.  As is well known, the spokes of a rotating wheel, and indeed the whole wheel, may be seen as stationary in a stroboscopic light.  For us it is important that at each point of the "stationary" wheel the maximal intensity of the light is constant, i.e. similar light pulses are repeated, at each point on the retina.

 It is obvious that rotation is not necessary for obtaining a stroboscopic picture, and only the frequency correlation is important.  One can replace a rotating object by a linear oscillating one, obtaining a linear stroboscopic picture.  It is also not necessary that the illuminated object have some visible elements of structure like the spokes.  The beam structure of the applied illumination is the alternative possibility.

 The following simple analogy helps one to understand the analysis of the next section. Assume that an angle-oscillating gun-machine is shutting onto a string while smoothly oscillating to and from, while the firing rate varies synchronously with the oscillation angle of the barrel.  After some time, the distribution of the hits on the string will obviously have some clear uneven pattern that indicate that some places of the string were hit by more bullets than the other places.

In the biological context, this analogy stresses that constant illumination can hardly be considered as a constant "signal", like, say, a constant pressure caused by a brick lying on a table. The density of the hits is increases as time passes, and the absorption of the light (the photons) also causes some *integral* effect in the photo-excited chemical processes and in the initiation of the nerve excitation, with possible over-accumulation of the products of the

Since now, we shall continue with only the geometric patterns of the averaged optical image on retina. Regarding these patterns, we shall show that for very different time shifts, two strongly distinct possibilities are obtained, and these also depend on the waveform of the light ripple. In terms of stroboscopic pictures, we have in one of the cases a simple stroboscopic picture, whose structure presents, however some strong local asymmetry meaning unequal excitation of some different groups of the involved nerve fibers. In the other case, a "doubled" stroboscopic picture in which the intensity of the pulses alternates between two different values is obtained, and we can assume that despite the physiological low-frequency filtration, this effect my be registered by the brain.

## 6. The tremor's influence on the shape of the image on the retina, and the role of the singularity of $\psi(t)$

Let us see now how the micro saccades and the waveform of $\psi(t)$ influence the shape of the image on the retina. Fig. 3 schematically shows the retina as a 1D "scanning system".

We start from the case of a purely sinusoidal ripple ("nonsingular case"). The role of the phase control is well seen in this case, and we shall see then the distinction with respect to the singular case, which is necessary for understanding the specificity of the fluorescent lighting compared with, say, LED's lighting.

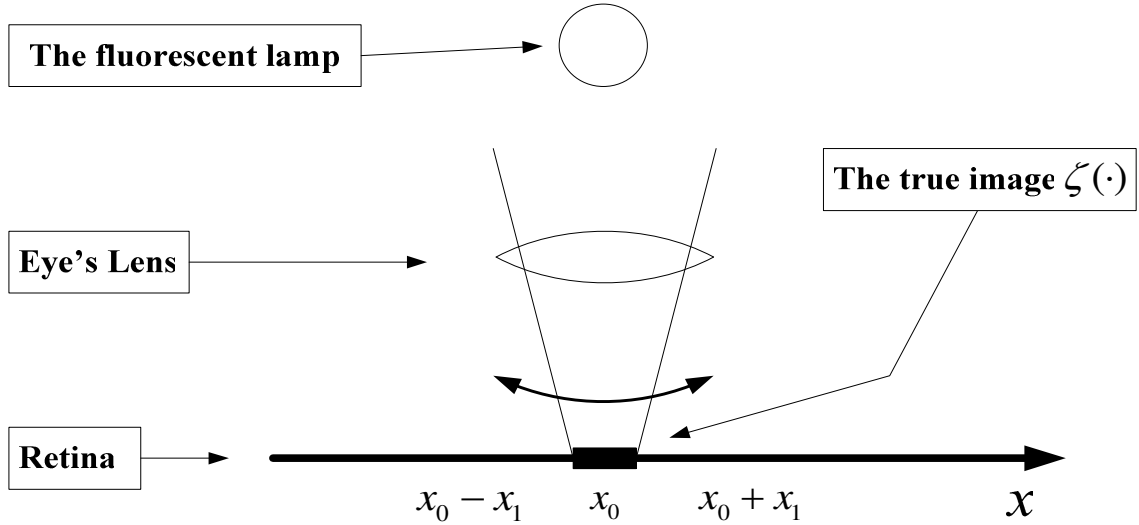

Figure 3 - The localized image on retina, and the micro saccadic movement (only such we consider here) that causes the 'true' image $\zeta(\cdot)$ to move on the retina. $x_1$ is the amplitude of the movement.

### 6.1. Nonsingular case

For the sinusoidal ripple,

$$\psi(t) = \Psi + B \sin 2\omega t , \qquad (6)$$

with a constant $0 < B < \Psi$, we consider a "1D retina" (or only the $x$-axis of a 2D retina). It is assumed that in its dependence on $x$ the image $\phi$ (or $A$) is a sharp pulse $\zeta(\cdot)$ ($\zeta_{max} = \zeta(0)$), that, *without the tremor*, is centered on a certain point $x_0$ on the retina. Thus, without the tremor, the received signal is directly proportional to $\psi(t) \cdot \zeta(x - x_0)$.

The tremor causes the position of the image on the $x$-axes to oscillate with some amplitude $x_1$. Assuming ideal synchronization between the ripple of $\psi(t)$ and the tremor, we have the position of the pulse to be $x_0 + x_1 \sin 2\omega t$, and the signal to be proportional to the expression

$$(\Psi + B \sin 2\omega t)\zeta\left(x - x_0 - x_1 \sin 2\omega t\right), \quad -x_1 \le x - x_0 \le x_1 . \qquad (7)$$

A more general 1D eye movement results in the factor $\zeta(x - \xi(t))$ with some $\xi(t)$. The condition $x - \xi(t) = const$ where the '*const*' (e.g., 0) belongs to the range of $x$ where $\zeta(\cdot)$ is nonzero, defines the movement of the modulated-in-time pulse. Differentiating the function $x(t)$ *defined by the equation* $x - \xi(t) = const$, we have the velocity of the movement of the pulse along the $x$-axis, $dx/dt$, to be $d\xi/dt$. For (7), this velocity is $2\omega x_1 \cos 2\omega t$.

The "sharpness" of $\zeta(\cdot)$ means that its width is much smaller than the scanning range $2x_1$, and thus we can speak about the appearance of the pulse "at some $x$", omitting the word "vicinity".

According to (7), the sharp image $\zeta(\cdot)$ arrives at a certain chosen point $x$, belonging to interval $[x_0 - x_1, x_0 + x_1]$ at the instants $t$ defined by the equality $x - x_0 - x_1 \sin 2\omega t = 0$, i.e.

$$\sin 2\omega t = (x - x_0)/x_1. \tag{8}$$

Using this in (7), we obtain

$$[\Psi + (B/x_1)(x - x_0)]\zeta(0), \qquad -x_1 \le x - x_0 \le x_1. \tag{9}$$

Thus, when the signal moves, because of the tremor, forward and back on the retina synchronously with its modulation in time, the *maximal* value of the signal, which the receptors at the vicinity of the point $x$ sense, depends on $x$ linearly. The pulse of the height (9) arrives at each $x$ twice per period of the modulation, once on its way forward, and once on its way back. Since the pulse periodically appears and disappears at each '$x$', the (essentially positive) signal depends, for each '$x$', on '$t$', but is smoothed as the time function by the inertia of the chemical processes in the receptors, i.e. by the delay in the reduction of the non-excited state when the light disappears in '$x$'. Because of this smoothing, the linear dependence of (9) on '$x$' may be seen as approximating the form of the actual retinal response/excitation to the flickering light, and we shall call such a dependence as *"image's shape on the retina"*, or "*image on the retina*". The role of the pulse $\zeta(\cdot)$ may be simply compared with edge of a pencil that draws this image.

Frequency synchronization, may be, in principle, with different time, or phase, shifts. Introducing into the tremor the phase shift of 180°, we obtain in (9) '-$B$' instead of $B$, which means similar slope of the "image" along the $x$-axes, but in the opposite direction. This is a simple case which should not influence the vision process.

Let us introduce into the tremor a 90° phase-shift. Then, the pulse position becomes $x_0 + x_1 \cos 2\omega t$. Now (compare to (8)), $\cos 2\omega t$ gives the range for $x$, and, considering that for a certain value of $\cos 2\omega t$ there are two values of $\sin 2\omega t$, we now obtain that on the way from $x_0 + x_1$ to $x_0 - x_1$ the maximal value of the pulse at $x$ is not (9), but

$$\left[\Psi + B\sqrt{1 - \left(\frac{x - x_o}{x_1}\right)^2}\right]\zeta(0), \quad -x_1 \le x - x_0 \le x_1, \tag{10}$$

and on the back way, i.e. from $x_0 - x_1$ to $x_0 + x_1$,

$$\left[ \Psi - B\sqrt{1 - \left(\frac{x - x_o}{x_1}\right)^2} \right] \zeta(0), \quad -x_1 \le x - x_0 \le x_1. \tag{11}$$

The non-zero difference

$$2B\zeta(0)\sqrt{1 - \left(\frac{x - x_o}{x_1}\right)^2}$$

between (10) and (11), whose largest value, $2B\zeta(0)$, is obtained at $x = x_0$, means that the envelope of the image is non-constant, in the interval $(x_0 - x_1, x_0 + x_1)$. We see that for the 90°-shift the "image on retina" is modulated in time with the basic frequency *2ω*, at each *x*.

Contrary to (9) which is odd with respect to $x_0$, functions (10) and (11) are even with respect to $x_0$. That the symmetries are very different should be taken into account when considering the signals within the fibers of the vision nerve. The fibers involved in the vision may be classified as related to spatial groups which are excited with different intensities.

Figure 4 draws the spatial functions (9,10,11), illustrating the above. Expressions (10) and (11) lead, respectively, to the upper and lower halves of an ellipse, while the straight line connecting the points $\left(x_0 - x_1, (\Psi - B)\zeta(0)\right)$ and $\left(x_0 + x_1, (\Psi + B)\zeta(0)\right)$ is described by expression (9).

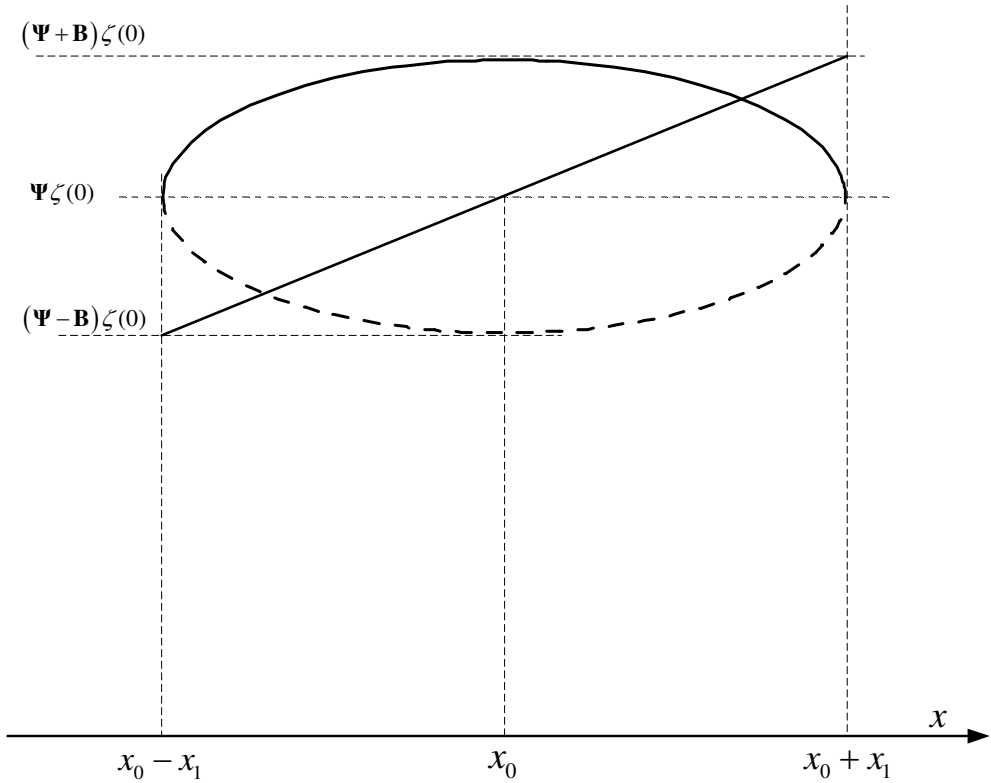

<u>Figure 4</u> - The different symmetries of the spatial distribution of the image intensity. The straight line segment for (9), and the very different ('elliptic') case of (10,11), for the movement of the pulse forward and back. In the 'elliptic' case, the intensity is modulated (oscillates) at each point between the values given by the continuous and the dashed lines. At $x = x_0$ the modulation is most strong.

Let us illustrate the above formulae argument by some numerical simulations, directly showing how $\zeta(\cdot)$ "draws the image".

The numerical model used for the generations of Figs. 5 and 7, included $\zeta(\cdot)$ as the cosine shaped, positive polarity pulse

$$\zeta(a) = \cos(2\pi \cdot 2500a)[u(a + 10^{-4}) - u(a - 10^{-4})]$$

where $u(\cdot)$ is the unit-step function (0 for negative and 1 for non-negative argument). Note that the exact shape of the sharp pulse is not seen in any of the figures, and actually is not so important for the sake of the illustration (e.g. narrow *rectangular* pulse gives a similar picture.) For the graphical illustration we used in (10,11) radial frequency $\omega = 2\pi 50 \text{ rad}/\sec$, $B = 2$, $\Psi = 5$, $x_0 = 0.01$, $x_1 = 3 \cdot 10^{-3}$.

The noted distinction between the cases of the zero and 90° phase shifts, is illustrated in Figs 5 (a) and (b) (as well as in Figs 7(a) and (b) below) in a "dynamic" manner. This is, actually, the evolution of the process in the $(x,t)$-plane. For each figure, one takes a point at the horizontal $x$-axis, and goes up on the vertical line, observing how the intensity (the blackness of the strip), at the chosen spatial point, is changed in time. Figure 5(a) gives a realization of (9), i.e. of the straight line in Fig. 4, and Fig. 5 (b) gives a realization of (10, 11), i.e. of the ellipse in Fig. 4.

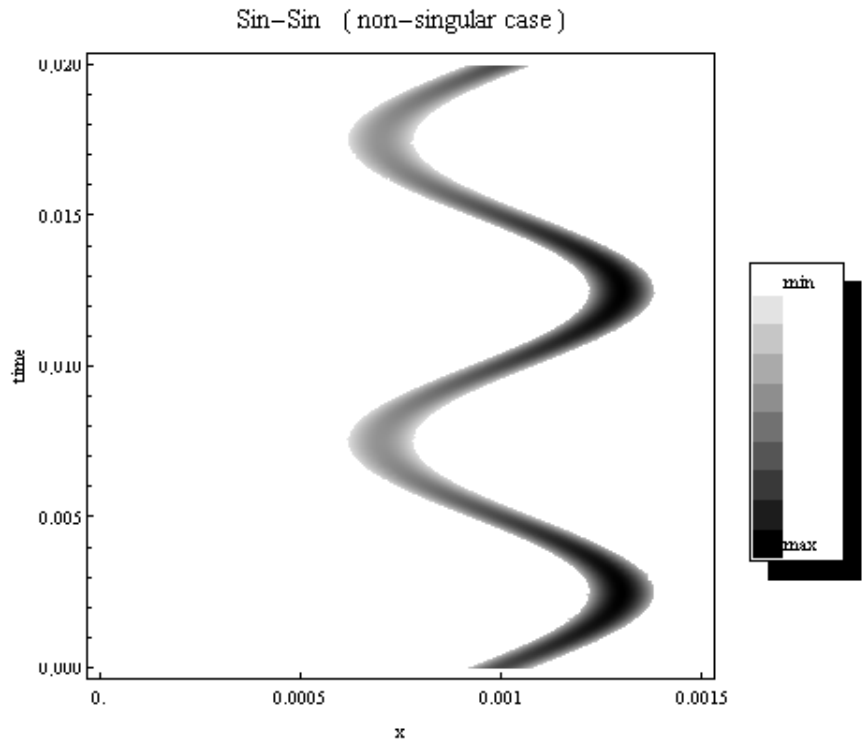

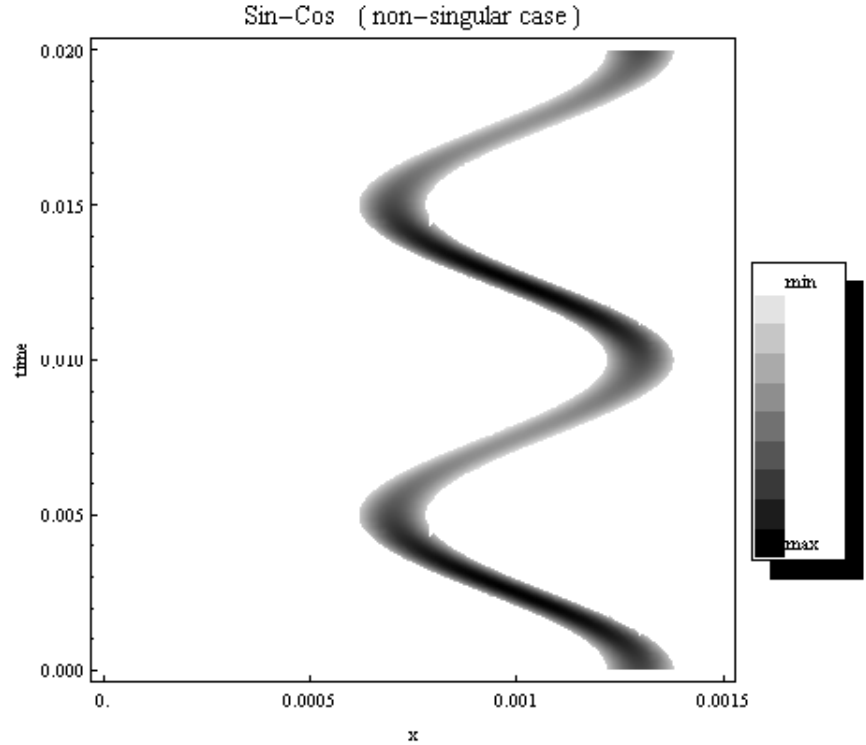

<u>Figure 5</u> (upper-(a); lower-(b)) - The demonstration, by simulation, of the influence of the mutual phase shift between the eye tremor and the light flickering, in the synchronous movement, for the case of sinusoidal modulation. Compare the distribution of the intensity (blackness of the strip), shown in this figure, with equations (9,10,11) or with the patterns of Fig. 4. (a) No phase shift ("sin-sin"), the "simple stroboscopic case", (b) 90° shift ("sin-cos"); the "doubled stroboscopic case".

The non-constancy, at each point, of the maximal value of the signal in the case of a 90°-shift states a noticeable distinction between this case and the case of a zero phase shift.

*Regarding the physiological processes in the retina per se*, one can conclude that if, in the case of 90°-shift, the oscillating signal makes the receptors' chemical operation easier, then the phase control should provide the 90°-shift.

### 6.2. The singular case

The important role of the phase (or, rather, *time*) shift also remains when the ripple is singular. However, the cases when the intensity of the 'image' is oscillating/modulated, or constant, are interchanged. For the singular model of $\psi(t)$ (compare with (1), noting that precise form of $i(t)$ is not given and is not very important here) we set $\psi(t) = \psi_{\min} + B|\sin \omega t|$, having instead of (7):

$$\left(\psi_{\min} + b\left|\sin \omega t\right|\right) \cdot \zeta\left(x - x_0 - x_1 \sin 2\omega t\right) . \qquad (12)$$

That as regards the cases of the modulated or constant "image", this situation is opposite to that of Fig. 5, may be very simply explained by the symmetry features (evenness, oddness) of the factors in (12) and also in (13). However, the following detailed considerations of this very interesting point are methodologically useful, as they allow one to consider cases of more complicated functions, not possessing such symmetry features.

Figure 6 graphically explains, according to (12), what is the intensity of the pulse when it comes to a point $x$. The graph of $\left|\sin \omega t\right|$ is drawn near the graph of $\sin 2\omega t$ crossed by the horizontal $x$-dependent level $(x - x_0)/x_1$, $-x_1 \leq x - x_0 \leq x_1$. The values of $\left|\sin \omega t\right|$, relevant to the intensity of $\psi(t)$, i.e. to the first factor in (12), are thus taken at the instants (the crossings of $\sin 2\omega t$ by the level) where the argument of the pulse-form $\zeta(\cdot)$ is zero, i.e. $\zeta(\cdot)$ is maximal.

One sees, regarding (12), that for $x > x_0$, and when the pulse is "coming" from $x_0 + x_1$ to $x_0$, $\zeta_{\max} = \zeta(0)$ is multiplied first by a relatively small factor, and then by a larger factor, at each point. For the backward direction, the same factors appear in time in the inversed order. For movement of the pulse in the region of $x < x_0$, the situation is also obvious from the figure.

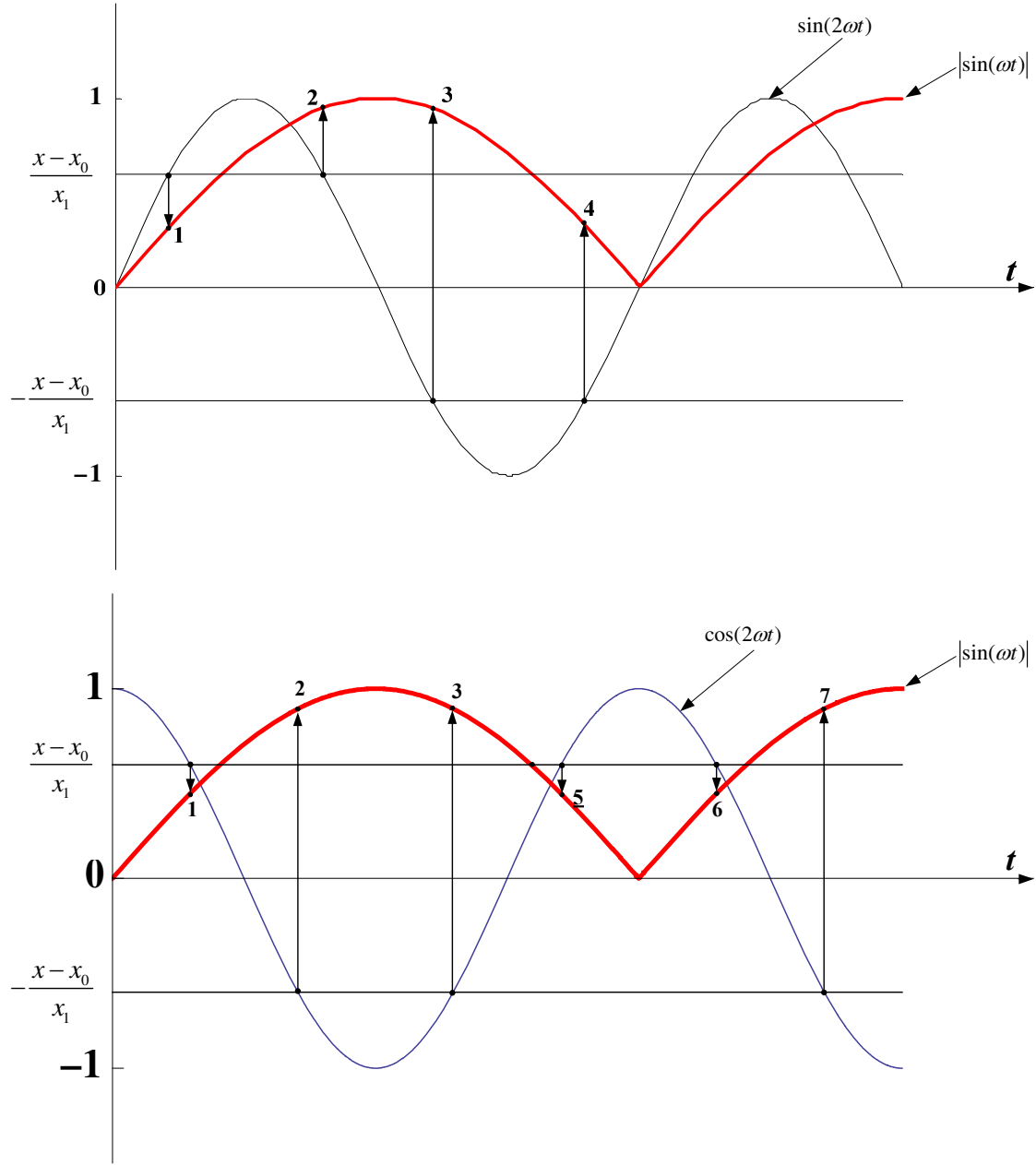

<u>Figure 6</u> - The drawing helping in obtaining the maximal values of the light intensity at $\forall x \in (x_0 - x_1,\ x_0 + x_1)$, in (12). Compare with Fig. 7(a) below. For $0 < \dfrac{x - x_0}{x_1} < 1$, points 1 and 2 give the values of $\left| \sin \omega t \right|$ at the instants when $\zeta(\cdot)$ possesses its maximal value $\zeta(0)$, and for $-1 < \dfrac{x - x_0}{x_1} < 0$, points 3 and 4 give such maximal values. We thus see what are the maximal light intensities at any relevant $x$, and obtain the "image on the retina" associated with the micro saccadic movement. Passing in this figure from $\sin 2\omega t$ to $\cos 2\omega t$ (the case of (13)), we obtain merging of points 1 and 2, and also 3 and 4, and these four points become two points, placed at the same height. This explains disappearance of the modulation of the "image" on the retina, which is seen also in Fig. 7(b).

A simple graphical consideration, similar to that of Fig. 6, shows that for a 90° shift in the tremor, i.e. for

$$\left(\psi_{min} + b\left|\sin\omega t\right|\right) \cdot \zeta(x - x_0 - x_1 \cos 2\omega t) \qquad (13)$$

we have, for every *x,* two pulses *of the same height*, per the period of the tremor/flickering.

Figures 7 (a) and (b) present, similar to Fig. 5, the "dynamic" simulation, illustrating the role of the phase shift for the case of the singular $\psi(t)$, confirming the above conclusions obtained using Fig. 6.

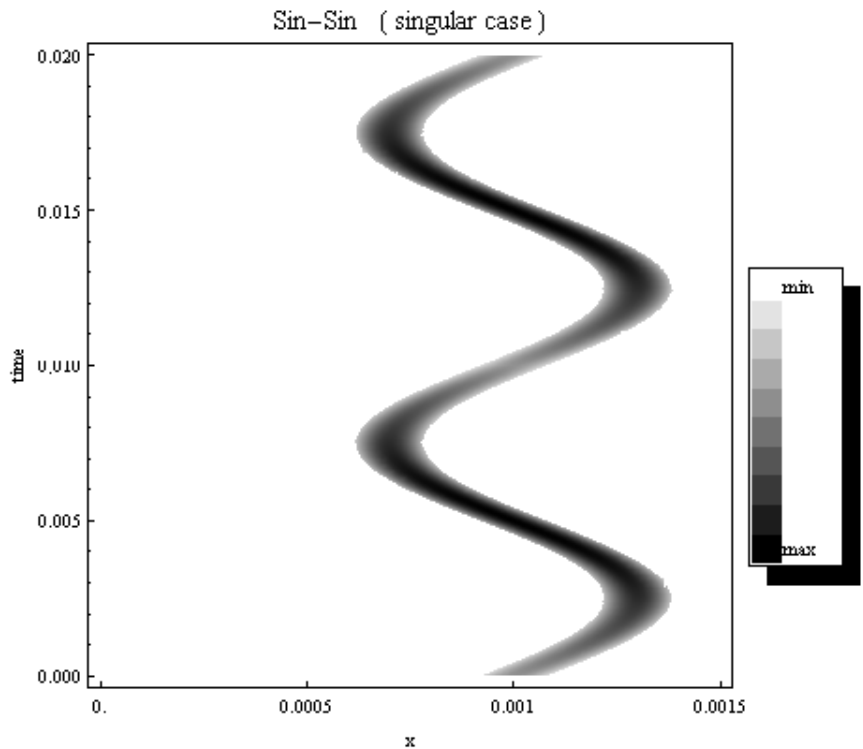

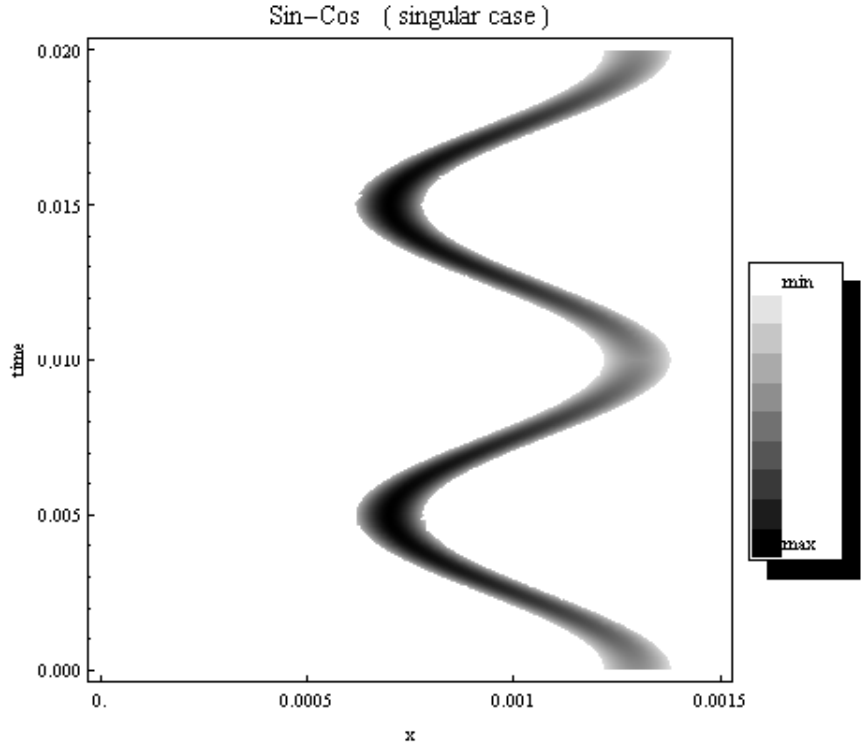

That is, the situation with respect to the case of non-singular $\psi(t)$ is opposite, and we see that whether of not the "image" on retina will be oscillating for a given time-shift depends on whether or not $\psi(t)$ is singular.

It is easy to see that for the singular $\psi(t)$ , the range of the intensities is somewhat narrower than for the nonsingular case, while, here also, the pulses arrive twice per period at each point.  However, now the *zero phase-shift* gives an oscillating maximal value in each point, while the 90° shift gives a constant maximal value.  One sees that the singularity of $\psi(t)$  is important for the analysis of the processes on the retina, i.e. degree of the difficulty for the brain to control the eye.

A final note relates to the quantitative side of the simulations.  In order to have a realistic comparison of the two singular cases with those nonsingular shown in Figs 5 (a) and (b), we consider that the limits of the light-intensity modulating factor in (9-11), $\Psi \pm B$, have to be similar to those for the singular case.  Thus, in order to obtain the limiting levels that, for the numerical values realized in Fig. 5, become in Fig. 4 $(5-2=3, 5+2=7)$, we use for the numerical model of (12) and (13) $\psi_{min} = 3$ and $b = 4$, while the values of $x_0$ and $x_1$ are taken identical to those used for Fig. 5.

## 7. On the role of the amplitude of the tremor

According to our argument, the brain controls, at least partly, the eye muscles' tremor, by controlling its phase with respect to the phase of $\psi(t)$. However the biological control cannot influence only the phase, and the amplitude of the tremor must also be influenced. The role of $x_1$ in (9) is worth considering. The larger $x_1$ is, the weaker are the variations of the maximal values of the signal appearing at a fixed distance from $x_0$, while close to $x_1$ the variations are always large. For autists who usually have some tendency to peripheral vision, an increase in $x_1$, that is, in amplitude of the tremor, may be a natural reaction.

Assuming now that the "pulse" $\zeta(\cdot)$ is *wider* than $2x_1$ (i.e. the lamp is relatively close to the eye), we observe that because of the similar (and high) intensity of the light emitted by each point of the tube's surface, the FL is seen (for any form of the tube) as 'flat' from any direction. Thus, despite the tremor, for a sufficiently wide shape of $\zeta(\cdot)$ (i.e. of the close tube), there is a continuous $x$-interval on the retina, narrower than the width of the shape, where the synchronously flickering signal, of similar intensity at each point, exists. That is, the micro-saccades stop to be effective from any point of view.

The necessity to decrease the average (5), i.e. the basic vision problem, should remain also for the case of the close lamp, but influence of the singularity of $\psi(t)$ should be different.

The assumption that the autist's reaction may increase $x_1$, means that the cases of distanced and not very distanced lamp may be made by the autists to differ not strongly. Comparative analysis of the reactions to the distanced and close lamps is one of the interesting points that may be suggested for experimental investigation.

## 8. An alternative point of view on the role of frequency correlation.

Since the existing state of physiological vision science does not allow one to say anything with certainty about of the brain's preference for images patterns of this or that kind, it is reasonable to also point out some other interpretations of the interfering role of high-frequency light flickering.  In essential contradistinction to the discussion in Section 3, the following argument uses that, as was argued in Section 3.2, the time averaging is not ideal over the inputs of the fibers of the optical nerve, and the brain detects, to a degree, the 100 Hz light signal.  However, the additional, not immediate point is that such a signal need not be caused only by a flickering light.

### 8.1.  On the role of the 100 Hz signal passing on to the brain

If the 100 Hz signal is registered by the brain, then the correlation between the flickering and the tremor may influence vision as follows.  When one looks, e.g., at a stationary border of black and white, the cortex receives a 100 Hz light-signal component *caused by the tremor*, i.e. the eye movement, *and such a signal would be a usual (unconscious) experience for the cortex, indicating that the vision process is in its normal state*.  Changes occurring in this usual signal because of the discussed correlation of $\psi(t)$ with the tremor, may be a reason for a problem in perceiving the visual information.  Modeling the scheme of Fig. 2 may be helpful for analysis of this point of view too.

The assumption that tremor normally (often) leads to a light-type signal of 100 Hz that is registered by the brain permits even a strongly distinct interpretation of the role of the frequency correlation.  If the brain needs the tremor only for indicating, by means of the returned 100Hz signal, that the vision process is normal, then when the 100 Hz *light-flickering* appears, causing a 100 Hz signal which is *more intense than usual* to enter the brain, the brain may lose any control of the tremor, and, together with this, the control of the low-frequency saccades, performed via the same muscles, which could be serious disturbance of the regular visual cortex activity.

Observe that also from this alternative point of view, the control of tremor exists, but now the role of the phase adjustment should not be so important.

### 8.2.  On the case when one's tremor frequency is not close to 100 Hz

Another case of the vision problem may be found in people, *whose eye-muscles' tremor frequency is not close to 100 Hz*.  In such humans, *the cortex is not adjusted to the 100 Hz signal*, and thus the ripple of $\psi(t)$ , -- even when associated with a very weak signal to the brain, -- could be *consciously* registered by the brain, as redundant and irritating information, irrelevant to the simple lighting purposes given to the lamp.

Since this reason is not associated with the intensity of the brain noise, the latter problem should be relevant even to otherwise absolutely normal humans.  Note also that the waveform of the ripple of $\psi(t)$ , as well as the kind of the flickering lamp, are not important here.  It would be very interesting regarding this outlook to check whether or not the frequency of the tremor is being changed in those humans in whom problematic perception of the flickering fluorescent light appears after traumatic accidents.

## 9.  Some open physical problems

Vision research is associated with many interesting biophysical problems, some of which are described, e.g., in [3-16].   Let us formulate, here also, some "macroscopic" physical questions/statements that arose during the discussion.   Certainly, some partial answers (especially to the first question below) could be given, but we would like to stress the importance of obtaining full answers.

1.  Constant illumination on surface of a body is associated with energy consumption by the surface, and then with energy propagation inside the depth of the body.  Are such consumption and propagation in the retina sufficiently effective in order to maintain a certain state of the surface retinal processes, so that a 'constant' signal on the retina is obtained?    Since for the inputs of the visual nerve fibers, the 'signal' is not given directly by the light, but by the products/effects of the chemical processes, -- if, for a constant illumination level, these products are accumulated on the retina, then *the constancy of the light intensity does underline{not} mean constancy of the signal* which is transferred to the brain.  Either this physiological signal is being increased in an integral manner, or, on the contrary, the accumulated excessive products of the photo-excited chemical process prevent the excitation of the nerve-impulses from being transferred through the fiber.  What precisely happens at the input of the fibers of the vision nerve when the fixation is too prolonged, i.e. when the intensity of the light signal is constant near the input, for a long time?

2.  In [8] direct proportionality of the velocity and amplitude of micro saccades is stated.  This is a typical property of *solitary waves* [34,35], and the nerve pulses controlling the tremor, which come to the eye muscles, should be of this type.  Indeed, the movement-reaction of the eye must reflect the frequency and intensity of the acting mechanical forces that, in their turn, reflect the features of the coming nerve-pulses.  An interesting point is that in some physical situations, the initiation of a "running pulse" of a proper solitary form may require [34,35] electrical pulses of a special form to be generated at the input of the propagation line.

3.  We know that saccades are associated with the necessity to allow relaxation of the chemical response, letting it to be restarted again and again.  However, both the slow saccades and/or the tremor may also influence the chemical reaction by another way, namely *by causing mechanical vibration* (*acceleration*) *of the retinal fiber*s.  If a mass performs oscillations with frequency $\omega$, then the acceleration is proportional to $\omega^2$ and the amplitude of the oscillations.  Thus, since the ratio of the frequencies is about 20-30, in order to cause the same acceleration of the retina's fibers, micro saccades can be of much smaller (about 1/600) amplitude than the low-frequency saccades, i.e. of order 1 angular second.   The micro saccades may even become the main factor in such a purely dynamic influence on the chemical processes, especially in view of our assumption about the possibility in autists that the amplitude of the tremor may be increased during the anxiety state.  Can such a dynamical influence be expressed in terms of the rate of the chemical reaction?

## 10.  Conclusions and final remarks

A phenomenological point of view, formulated in some system terms, on the influence of light-flicker on human vision, is suggested.   The coincidence of the basic frequencies of the ripple of $\psi(t)$ and the eye's tremor is shown to be important, and our main assumption is that "brain noise" in autists prevents the necessary control of the tremor from being well performed, this being an important part of the vision problem.  For seeing the role of the tremor, it is sufficient (Section 6) to use a 1D model of the retina.   It needs to be stressed that the initial part of the problem related to the necessity for the brain to manage with the specific (in either the "resonant" or "stroboscopic" interpretations) forms of the patterns, -- relates to all humans, not just to autists or any other specific group.

In deriving our final conclusions, we considered that some signals are consciously registered by the brain, and some unconsciously, and that on the subliminal level, relatively high frequencies can be registered.   The simultaneous existence of the conscious and unconscious registrations is seen to be important for the brain operation of both autists and normal humans. (See also [22,23,37-38], and, in particular, [37] for a relevant example, associated with vision, of regulation of the behavior.)

The concept of the "scanning" of the information field on the retina should be completed by biological information about the order in which the signals in different fibers of the optic nerve are processed. The time-constant of deactivation of the receptors was considered only via its qualitative influence. The micro-saccades were discussed independently of the 3-5 Hz saccades and blinking, though one sees that for the latter, there are very different time scales, anyway. Any presumed difficulty for *both eyes* to simultaneously perform the needed control of the tremor, could not be considered here. Despite these limitations, the system-theoretic point of view on the very difficult vision problem/process creates a certain conceptual framework and formulates certain points for development.

Even though the physiology of human vision is not very close (perhaps, as a bird to an airplane) to the technological methods of *vision chips* [24-33], study of the vision *disturbance* problem should be essentially helped by using the ideas and methods of artificial vision science, because these methods allow one to *separately model* the mutually influencing factors in the cortex-eye nerve connection, which is impossible in the biological study. The control loop shown in Fig. 2 may be modeled in a chip, and imitation by the chip of the autistic reaction to changes in light intensity and the level of the brain noise can allow one to predict, in different situations, the behavior of an autist who by himself is not suitable for the "laboratory study". Thus, a teacher might check whether or not the illumination in an art-museum would permit visits by autists, or choose a proper place for an autistic student in a lecture hall, etc.. If the properties of such a chip/device could be adjusted to the characteristics of a particular autistic person, -- this would be a very important step!

## References

[1]   K.D. O'Leary, A. Rosenbaum, P.C. Huges, "Fluorescent lighting: a purposed source of hyperactive behavior", Journal of Abnormal Child Psychology, vol. 6 (1978), 285-289.  See also **http:// www. autism.org/html** for the specific autistic behavior problem.

[2]   J.A. Veitch, S..L. McColl, "A critical examination of perceptual and cognitive effects attributed to full-spectrum fluorescent lighting", Ergonomics, 2001, vol. 44, no.3, 255-279.

[3]  S. Martinez-Conde, S.L. Macknik, and D. H. Hubel, "The role of fixation eye movements in visual perception", Nature Reviews | Neuroscience, vol. 5 (March 2004), pp. 229-240.

[4] D.C. West, P.R. Boyce, "The effect of flicker on eye movement", Vision Research, Vol. 8 (1968), 171-192.

[4a]  A Spauchus, J. Marsden, D.M. Halliday. J.R. Rosenberg, P. Brown, "The origin of ocular microtremor in man", Exp. Brain Res. (1999) 126: 556-562.

[4b]   C.Bolger, S. Bojanic, N.S. Sheahan, D. Coakley, J.F. Malone, "Dominant frequency content of ocular microtremor from normal subject", Vision Research 39 (1999), 1911-1915.

[5] I. Murakami, P. Cavanagh, "Visual jitter: evidence for visual-motion-based compensation of retinal slip due to small eye movements …", Vision Research, vol. 41 (2001), 173-186.

[6] S.L. Macknik, B.D. Fisher, and B. Bridgeman, "Flicker distorts vision space constancy", …", Vision Res., vol. 31 (1991), 2057-2064.

[7] Eizenman at al, "Power spectra for ocular drift and tremor", Vision Research,  vol. 25 (1985), 1635-1640.

[8] B.L. Zuber, L. Stark, "Microsaccades and the velocity-amplitude relationship for saccadic eye movement", Science, vol. 150 (1965), pp.1459-1460.

[9] S.H. Bartley, *"Vision"*, Hafner, New York, 1963

[10] M.H. Pirenne, *"Vision and the eye"*, Chapman and Hall, London, 1971

[11] P.A. Buser, *"Vision"*,  Cambridge, Mass. MIT Press, 1992.

[12]  B. Cohen, I. Bodis-Vollner (Eds),*"Vision and the brain"*, Raven Press, New York, 1990.

[13] R.H.S. Carpenter (Ed.), *"Eye Movements"*, Houndmills, UK: Macmillan Press, 1991.

[14] D. Marr, *"Vision"*, W.H. Freeman, San Francisco, 1982.

[15] D. Regan, *"Spatial Vision"*, Houndmills, UK: Macmillan Press, 1991.

[16]  J.J. Kulikowski, V. Walsh and I.J. Murray, *"Limits of Vision"*, Houndmills, UK: Macmillan Press, 1991.

[17]  S.  Pinker, "*How the mind works*", W.W.  Norton, New York, 2004.

[18]  W.  Elenbaas, "*Fluorescent lamps and lighting*", Philips Technical Library, New York, 1962.

[19] M.V.  Kostic, F.V.  Topalis, "Survey for the theoretical methods for the interior lighting calculations", International Journal of Lighting Research and Technology, The Chartered Institution of Building Services Engineers, London, UK, CIBSE Series B, vol.  30, no.  4, pp.  151-157, 1998.

[20]  E. Gluskin, I. Kateri, F.V. Topalis, N.  Bisketzis, "The instantaneous light-intensity function of a fluorescent lamp", Phys. Lett A, 353 (2006), 355-363.

[21] E.  Gluskin, "The fluorescent lamp circuit", IEEE Transactions on Circuits and Systems, Pt.-I: Fund.  Theory and Appl., 46 (no.  5) (May 1999) pp 529-544.

[22] I.P.  Pavlov, "*Conditioned Reflexes*", Dover, New York, 2003 (Originally, Oxford University Press 1927.)

[23] I.P.  Pavlov, "*Lectures on Conditioned Reflexes: Twenty-five Years of Objective Study of the higher Nervous Activity Behavior of Animals*", Pinter, New York, 1980.

[24] E.  Gluskin, "Spatial filtering through elementary examples", European Journal of Physics, vol.  25 no.  3 (May 2004), 419-428.

[25]  A.E.  Gygi, G.S.  Moschytz, "Low-pass filter effect in the measurement of surface EMG", Proceedings Tenth IEEE Symp.  on Comp.  Based Medical Systems, (Cat.  No. 97CB36083) IEEE Comput.  Soc.  Press, Los Alamitos, CA, 1997.

[26]  T.  Yagi, Interaction between the soma and the axon terminal of retina horizontal cells in *cyprinus carpio*", J.  Physiol., vol.  375 (1986), pp.  121-135.

[27] B.E.  Shi, L.O.  Chua, "Resistive grid image filtering: input/output analysis via the CNN framework", IEEE Trans.  on CAS – Pt.  I, vol.  39, no.7 (Jul.  1992), pp.  531-547.

[28]  C.  Koch, H.  Li, "*Vision Chips*," Los Alamitos, CA: IEEE Computer Society Press, 1995.

[29] C.  Mead, "*Analog VLSI and Neural Systems*", Addison-Wesley, Reading, MA.  (1989).

[30] T. Delbrück, "Silicon retina with correlation-based, velocity-tuned pixels", IEEE Transactions on Neural Networks, vol. 4, no. 3 (1993), pp. 529-541. (Originally Report NN 127, Caltech, Pasadena CA 91125.)

[31] S.-C. Liu, J. Kramer, G. Indivery, T. Delbrück, T. Burg, R. Douglas, "Orientation-selective a VLSI spiking neurons", Neural Networks, 14 (2001), 629-643.

[32] B. Shi, "A low-power orientation-selective vision sensor", IEEE Transactions on Circuits and Systems, 47 (5) (2000), 435-440.

[33] K. A. Boahen, A. G. Andreou, "A contrast sensitive silicon with reciprocal synapses", Advances in Neural Information Processing Systems, 4 (1992), 764-772.

[34] A. Scott, F.Y.F. Chu and D.W. McLaughin, "The soliton – a new concept in applied science", Proc. IEEE, vol. 61 (1973), pp. 1443-14483.

[35] E. Gluskin, "Nonlinear systems: between a law and a definition", Reports on Progress in Physics A, vol. 60, no.10 (1997), 1063-1112.

[36] W. Heisenberg, "*Tail und das Ganze"* (*Part and Whole*), Piper, Munich, 1972.

[37] Y.S. Bonneh, A. Cooperman and D. Sagi, "Motion-induced blindness in normal observers", Nature, Vol. 411, 14 June 2001, pp. 798-801.

[38] R. Ben-Yishai, R. L. Bar-Or and H. Simpolinsky, "Theory of orientation tuning in visual cortex", Proc. Natl. Acad. Sci., USA, 92 (9) (1995), 3844-3848.